\documentclass[nofootinbib,showkeys,preprint,aps]{revtex4-1}
\usepackage{color}
\usepackage{amssymb} 
\usepackage{amsmath,amsfonts}
\usepackage{graphicx}
\usepackage{braket} 
\usepackage{hyperref}
\usepackage{subfigure}
\usepackage{upgreek}

\begin{document}

\title{Holographic Magnetic Susceptibility}

\author{Lei Yin\footnote{Corresponding to Lei Yin and Defu Hou }}
\affiliation{Institute of Quantum Matter , \\ School of Physics and Telecommunication Engineering, South China Normal University ,\\Guangzhou, 510006 , China.}
\email{yinlei@m.scnu.edu.cn}

\author{Hai-cang Ren}
\affiliation{Physics Department, The Rockefeller University,1230 York Avenue, New York, 10021-6399, U.S.A.\\ Central China Normal University,  Wuhan,  430079, China.}
\email{ren@mail.rockefeller.edu}

\author{Defu Hou}
\affiliation{Institute of Particle Physics and Key Laboratory of Quark and Lepton Physics (MOS) , Central China Normal University, \\ Wuhan,  430079, China.}
\email{houdf@mail.ccnu.edu.cn}

\begin{abstract}
 The (2+1)-dimensional static magnetic susceptibility in strong-coupling is studied via a Reissner-Nordstr\"{o}m-AdS geometry. The analyticity of the susceptibility on the complex momentum $\mathfrak{q}$-plane in relation
to the Friedel-like oscillation in coordinate space is explored.
In contrast to the branch-cuts crossing the real momentum-axis for a Fermi liquid, we prove that the holographic magnetic susceptibility remains an analytic function of the complex momentum around the real axis in the limit of zero temperature.  At zero temperature, we located analytically two pairs of branch-cuts that are parallel to the imaginary momentum-axis for large $|\text{Im}\ \mathfrak{q}|$ but become warped with the end-points keeping away from the real and imaginary momentum-axes.  We conclude that these branch-cuts give rise to the exponential decay behaviour of Friedel-like oscillation of magnetic susceptibility in coordinate space. We also derived the analytical forms of the susceptibility in large and small-momentum, respectively.
\end{abstract}


\maketitle

\section{Introduction}
\label{sec:introduction}
 
Strongly correlated electronic systems, such as the high temperature superconductors or graphene, are characterized by a spectrum of novel static and
transport phenomena that cannot be explained by the traditional Fermi liquid theory of Landau and are difficult to explore with ordinary field theoretic techniques. The perturbative expansion or
mean field approximation becomes unreliable, especially in lower dimensions, and the first principle numerical simulation is hindered by the fermion sign problem.
The holographic theory \cite{Witten1998b,Maldacena1998,Aharony1999,Klebanov1999,witten2001multi,Gubser1998} built on the conjectured gauge/gravity duality is expected to shed some lights on the non-perturbative physics and to reveal some generic properties pertaining
to a strongly-coupled system\cite{Casalderrey-Solana2006,Zaanen,Hartnoll2016}, such as a non-Fermi liquid\cite{Cubrovic2009,Faulkner2011,Liu2011,Faulknerab,Iqbal}.  According to the holographic dictionary, the classical solution of the gravity-matter
system in an asymptotically AdS space-time with a black hole is linked to the thermodynamics of a strongly coupled quantum field theory on the AdS boundary\cite{Witten1998}. In particular, the linearized solutions of the former generate various
two-point correlation functions of the latter \cite{Witten1998b,Gubser1998,Son2002a},  and the photon polarization tensor to be investigated in this work is one of them.
  
The general structure of the polarization tensor in energy-momentum representation, dictated by the current conservation, is given by
\begin{eqnarray}
  \Pi_{ij}(\vec q,\omega) &=& \chi(\omega,q)(q^2\delta_{ij}-q_iq_j)+\omega^2\alpha(\omega,q)\frac{q_iq_j}{q^2}\nonumber\\
  \Pi_{0j}(\vec q,\omega) &=& \Pi_{j0}(\vec q,\omega)=\omega\alpha(\omega,q)q_j\nonumber\\
  \Pi_{00}(\vec q,\omega) &=& q^2\alpha(\omega,q) \quad ,
\label{polarization}
\end{eqnarray}
with the transverse and longitudinal form factors, $\chi(\omega,q)$ and $\alpha(\omega,q)$, representing the magnetic susceptibility and electric polarizability,
respectively. Both variables $\omega$ and $q$ in $\chi(q,\omega)$ and $\alpha(q,\omega)$ can be continuated to the complex
planes. The singularities on the $\omega$-plane reflect the excitation spectrum, while the singularities on the complex $q$-plane give rise to the
Debye-like screening and Friedel-like oscillation in coordinate space. The analyticity of $\chi(\omega,q)$ and $\alpha(\omega,q)$ in weak coupling is
well known. For the (2+1)-dimensional static polarization tensor considered in this paper, the one-loop calculation of $\chi(0,q)$ and $\alpha(0,q)$ for a spinor QED reveals
two lines of square root branch points located at \cite{Yin2016}
\begin{equation}
    q = \pm 2 \big[ \mu + \mathrm{i} \pi T (2n + 1) \big]   \\
  \qquad n \in \mathbb{N} \; ,   \label{eq:1}
\end{equation}
with $T$ the temperature and $\mu$ the chemical potential. In the zero temperature limit $T\to 0$, these singularities merge into two cuts with $\rm{Re} \, q=\pm2\mu$ across the
real axis, which results in a discontinuity in the derivative of $\chi(0,q)$ and $\alpha(0,q)$ at $q=\pm 2\mu$ and the Friedel oscillation in coordinate
space with the amplitude decaying according to a power law.

In strong coupling, the holographic $\chi(\omega , q)$ and $\alpha(\omega ,q)$ extracted from different bulk geometries along with their analyticity have been discussed
extensively in the literature, such as Ref.\cite{Hartnoll2008,Hartnoll2008a,Hartnoll2009, Faulkner, Donos2014}  for $q = 0$, and Ref.\cite{Anantua2013,Hartnoll} for $\omega =0$.  (For more details, see Ref.\cite{Zaanen} and the references therein). In this paper, we shall focus on the momentum analyticity of the holographic polarization tensor from a Reissner-Nordstr\"{o}m-AdS
geometry. In the same system, Ref.\cite{Edalati2010e} studied the conductivity via the small frequency expansion in the IR limit of CFT, finding that the conductivity at zero-momentum scales as $\omega^2$ in $\omega \to 0$. For the $\chi(0,q)$ and $\alpha(0,q)$ extracted from the Schwarzschild-AdS geometry (corresponding to zero chemical potential), it was shown in Ref.\cite{Hou2010a} that all of
singularities on the q-plane are poles located along the imaginary momentum-axis. A similar result was obtained by a study on the probe D3/D5 system at a nonzero density in Ref.\cite{Anantua2013} and the authors revealed that such poles at the purely imaginary momentum screen exponentially a point charge in the medium  and do not cause  Friedel-like oscillation. Then came the work by Blake et.al.\cite{Blake2015c},
who solved the Einstein-Maxwell equations numerically for the gauge field and metric tensor fluctuations in the Reissner-Nordstr\"{o}m-AdS background with a complex momentum and found two lines of poles of $\alpha(0,q)$ whose locations tend to be parallel to the imaginary $q$-axis for large $|{\rm Im}q|$ and bend toward the imaginary axis at lower $|{\rm Im}q|$. Their numerical solution also indicates an exponentially decaying Friedel-like oscillation behavior even at zero temperature. In our previous works
\cite{Yin2016} and \cite{Yin2017}, we were able to prove that both $\chi(0,q)$ and $\alpha(0,q)$ extracted from the non-extremal Reissner-Nordstr\"{o}m-AdS geometry are meromorphic functions and to
locate their poles analytically for large $|\rm{Im}\, q|$ via WKB solution of the Einstein-Maxwell equations. The asymptotic distribution of the poles is given by
\begin{align}
  q \simeq  \mu  \bigg[ \pm  w  \pm \mathrm{i} \frac{\pi}{Q L_1} | n - \frac{1}{4} |  \bigg] ,
  \label{eq:48}
\end{align}
with the integer $n \gg 1$, where $L_1$ and $L_2$ are two elliptic integrals dependent of the temperature $T$, defined in Eqn.(\ref{eq:47}) of Appendix A. As the temperature $T\to 0$, the distance
between adjacent poles, $\frac{\pi}{QL_1}\sim \big(\log \frac{T}{\mu} \big)^{-1}\to 0$ and the poles merge into two pairs of cuts, parallel to the imaginary axis but at much slower rate
than the weak coupling case.
For $\alpha(0, q)$, the asymptotic locations (\ref{eq:48}) match well with the numerical result in Ref. \cite{Blake2015c} even with a moderate $\rm{Im}\, q$. Unfortunately,
the condition for the WKB prevented us from making any rigorous statements regarding the distribution of these poles near the real momentum axis, which may be more relevant to experimental observations.

This work is a continuation of Refs.\cite{Yin2016}. Different strategies are employed here to explore the analyticity of the holographic magnetic susceptibility
$\chi(0,q)$ in the complex $q$-plane, especially at zero temperature where the RN black hole becomes extremal. Through the series solution of the Heun equation
involved, we show that the complex poles of $\chi(0,q)$ discussed in \cite{Yin2016} merge into four branch cuts of square root type at zero temperature, whose trajectories are located analytically. Coming from the infinity, these cuts are nearly parallel to the imaginary axis for large $|\rm{Im}\, q|$, in agreement with the WKB approximation, bending towards
the imaginary axis for lower $|\rm{Im}\, q|$ and terminating at respective branch points with $|\rm{Im}\, q|\neq 0$ and $|\rm{Re}\, q|\neq 0$, without crossing either the real or imaginary axes
on their paths. Through a relation between the Einstein-Maxwell equations and the eigenvalue problem of an one-dimensional Schr\"odinger equation, we prove that $\chi(0,q)$ is an analytic function for any finite real $q$ at any temperature, which excludes
any oscillatory behavior caused by singularities on the real axis.
Consequently,the magnetic susceptibility manifests a Friedel-like oscillation in coordinate space which decays exponentially even down to zero temperature.

The analytic technique employed in this work is not yet to be generalized to the case of electrical polarization, $\alpha(0,q)$ in order to extend the result of Ref.\cite{Yin2017} to zero
temperature, in which case the Einstein-Maxwell equations involved are far more
complicated. We hope to report our progress along this line in near future.

The paper is organized as follows: In Sec.~\ref{sec:hologr-model-magn}, we formulate the holographic magnetic susceptibility dual to an Einstein-Maxwell system in
the background of a Reissner-Nordstr\"{o}m blackhole with an asymptotically Anti-de Sitter boundary. The analyticity of the magnetic susceptibility is explored in Sec.~\ref{sec:magn-susc-gaug}. The asymptotic forms for  small and large complex momenta are derived in Sec. \ref{sec:magn-susc-at},    Sec.~\ref{sec:disc-concl-} concludes the paper.

\section{Holographic Model for Magnetic Susceptibility }
\label{sec:hologr-model-magn}

\subsection{Background Solutions and Fluctuations in D= 2+1 space-time }
  \label{sec:class-solut-fluct}

  According to the holographic principle,  the generating functional of correlators of a strongly-coupled quantum field theory(QFT) defined in space-time $\mathcal{S}$ is associated with the partition function of
  a classical gravity-matter theory in a bulk bounded by $\mathcal{S}$. This relation, as was formulated by Gubser-Klebanov-Polyakov and Witten (GKPW)\cite{Witten1998b,Maldacena1998,Gubser1998,Aharony1999,Klebanov1999,witten2001multi}, is:
\begin{align}
  Z_\text{QFT} [ \mathring{\phi}_i] = Z_\text{Grav.}[ \mathring{\phi}_i] \label{eq:31}
\end{align}
where
\begin{align}
  Z_\text{QFT} [\mathring{\phi}_i] = \langle e^{\mathrm{i} \sum_i \int \ \mathrm{d}^dx \mathring{\phi}_i \mathcal{O}_i }  \rangle  ;
\quad Z_\text{Grav.}[\mathring{\phi}_i]  =  e^{\mathrm{i} S[ \phi_i^\star \to \mathring{\phi}_i ]} \, ,
\end{align}
with $\phi_i$ the bulk fields of the gravity-matter system and $\phi_i^\star$ the classical solutions whose boundary value $\mathring{\phi}_i$ representing the source for $Z_\text{QFT}$, conjugate to the operators $\mathcal{O}_i$.

The bulk action of the gravity-matter system considered in this work reads

\begin{align}
  S = \int \ \mathrm{d}^4 x \, \sqrt{-g}\; \bigg[  G_4\, \big( R - 2 \Lambda \big)-  K_4\, \big( F_{\mu \nu} F^{\mu \nu} \big)  \bigg] \, ,
  \label{eq:2}
\end{align}
where $R$ is the scalar curvature corresponding to the metric tensor $g_{\mu \nu}$, $\Lambda$ is the negative cosmological constant, in $D=3+1$ dimensional AdS space-time, $\Lambda = - \frac{3}{L^2}$,
$L$ is the AdS radius, and $F_{\mu\nu}$ is the electromagnetic tensor, $F_{\mu\nu} = \partial_\mu A_\nu - \partial_\nu A_\mu$,
corresponding to the gauge potential $A_\mu$. The mass dimension of the gauge potential is $[A_\mu]=1$ and that of the coupling constant $G_4$ is $[G_4]=2$. The coupling constant $K_4$ is thereby
dimensionless, $[K_4] = 0$.

The background solution $(\bar{A}_\mu \, , \, \bar{g}_{\mu\nu})$ of the Einstein-Maxwell equation dictated by the action ~(\ref{eq:2}) consists of the Reissner-Nordstr\"{o}m-AdS metric
\begin{equation}
  \ \mathrm{d} s^2=\bar g_{\mu\nu} \ \mathrm{d}x^{\mu}\mathrm{d}x^{\mu} =  \left(\frac{L}{z_+ u}\right)^2 \left(-f(u) \ \mathrm{d}t^2 + \frac{z_+^2}{f(u)} \ \mathrm{d}u^2 +\ \mathrm{d}x^2 +\ \mathrm{d}y^2 \right)
  \label{eq:33}
\end{equation}
and the gauge potential
\begin{equation}
  \bar A = \bar A_t \ \mathrm{d}t = \frac{Q}{z_+} ( 1 - u ) \ \mathrm{d}t \, ,
  \label{eq:34}
\end{equation}
where the metric function
\begin{equation}
  f = 1 - (1+Q^2)u^3 + Q^2 u^4  \, ,
  \label{eq:35}
\end{equation}
with the horizon of this Reissner-Nordstr\"{o}m black hole at $u=1$ and the AdS boundary at $u=0$. The chemical potential $\mu$ of the boundary field theory is related to the dimensionless charge of the black hole $Q$ via $\mu=Q/z_+$. The Hawking temperature $T$ in terms of the chemical potential and the charges reads
\begin{equation}
T=\frac{\mu}{4\pi}\frac{3-Q^2}{Q} \;  ,
\end{equation}
which corresponds to the background temperature of the boundary field theory, and $Q\in [0,3]$ and $L \sqrt{G_4/K_4}$ is re-scaled to 1. At $Q^2=3$,
it represents the zero-temperature limit with the extremal metric function $f_0 = (u-1)^2(3u^2 + 2u +1)$
holding a double zero at the horizon.

Introducing the metric and gauge potential fluctuations $(h_{\mu\nu}, a_\mu)$ via
\begin{equation}
  g_{\mu\nu}(X,u)=\bar g_{\mu\nu}(u)+h_{\mu\nu}(X,u) \quad ;\quad A_\mu(X,u)=\bar A_\mu(u) + a_\mu(X,u) \, ,
\end{equation}
with $X=(t,x,y)$, a nontrivial solution of the Einstein-Maxwell equations for $(h_{\mu\nu},a_\mu)$ is driven by their nontrivial values at the AdS boundary, $u=0$.
In terms of such a solution, the action
becomes a functional of the boundary values $h_{\mu\nu}(x,0)$ and $a_\mu(x,0)$ and the coefficients of the power series expansion of this on-shell action in
$h_{\mu\nu}(x,0)$ and $a_\mu(x,0)$ give rise to various correlation functions of the strongly interacting boundary field theory in accordance with the GKPW formula (\ref{eq:31}) .
The holographic counterpart of the polarization tensor (\ref{polarization}) corresponds
to the quadratic term in $a_\mu(x,0)$, hence the linearized Einstein-Maxwell equations thereby suffice for our purpose.

Owing to the homogeneity with respect to the boundary coordinates, $(x,y,t)$, the linearized Einstein-Maxwell equations can be Fourier transformed into the frequency-momentum space for
\begin{equation}
a_\mu(P ; u)=\int d^3X e^{-\mathrm{i} P\cdot X}a_\mu(X,u) \quad ; \quad h_{\mu\nu}(P ; u)=\int d^3X e^{-\mathrm{i} P \cdot X}h_{\mu\nu}(X,u) \, ,
\end{equation}
with $P=(\omega,q_x,q_y)$. Aligning the spatial momentum $\vec q$ along the $x$-axis, the linearized equations
can be decomposed into two decoupled subsets according to the parity under the transformation $y \to -y$ \cite{Edalati2010a}, i. e.
\begin{align}
  \textit{Even Parity}: \;\;  \{h_{\; t}^t , h_{\; x}^x, h_{\; y}^y, a_t \} \;  \text{and}  \; \{ h_{\; t}^x , a_x \}   ; \quad
  \textit{Odd Parity}: \;\;  \{h_{\; t}^y , a_y \} \; \text{and} \;  \{ h_{\; y}^x  \}
\end{align}
In the static limit ($\omega = 0$), each group of Einstein-Maxwell equations are further decoupled into the two subsets, denoted by the curly brackets above.
The electric component is extracted from the even-parity group, while the magnetic component from the odd-parity one, respectively.
The two coupled equations responsible to the static magnetic susceptibility read
\begin{align}
  a_y'' + \frac{ f' }{ f } a_y' - \frac{ Q^2 k^2 }{ f } a_y - \frac{ \mu }{ f } h'^y_{\; t} &= 0 \label{eq:49} \\
  h''^y_{\;t}- \frac{ 2 }{ u } h'^y_{\; t} - \frac{ Q^2 k^2 }{ f } h_{\;t}^y - 4 \frac{ Q^2 }{ \mu } u^2 a_y' &= 0  \label{eq:51} \; ,
\end{align}
where $Z = \frac{3}{4}\left( 1 + Q^{-2} \right)$, and we have introduced a dimensionless momentum $\mathfrak{q}\equiv q/\mu$ and the dimensionless modified momentum $k$ is defined by
\begin{align}
  k \equiv \sqrt{ \mathfrak{q}^2 + Z^2 } \; .
  \label{eq:60}
\end{align}
For the full set of Einstein-Maxwell equations in terms of our notations, see \cite{Yin2016,Yin2017}.
The static magnetic susceptibility at a temperature $T$ is given by
\begin{align}
  \upchi(q|T)\equiv\lim_{\omega \to 0} \chi(\omega,q)  = \frac{\mathcal{C}_{yy} }{q^2}   \label{eq:32} \, ,
\end{align}
where $\mathcal{C}_{yy}$ is the coefficient of $|a_y(P ; 0)|^2$ in the on-shell action with $P=(0,q,0)$, following the GKPW formulation (\ref{eq:31}) and Ref.\cite{Son2002a}, and it is dependent of the momentum $q$, temperature $T$ and chemical potential $\mu$ of the system.

In terms of the static solution of the linearized Einstein-Maxwell equations for $(a_y, h^t_y)$ that are regular at the horizon $u=1$ and subject to the boundary condition $h^t_y=0$ at $u=0$
(in order for extracting the polarization tensor only), the on-shell action becomes
\begin{align}
  S_{_\text{EM}} = - 2 K_4 \int \ \mathrm{d}^3x \bigg[ \sqrt{-g}  \bar g^{uu} \bar g^{\alpha \beta} \, a'_\alpha a_\beta  \bigg]\bigg|_{u=0} \, ,
  \label{eq:61}
\end{align}
where the prime refers to the derivative with respect to the dimensionless radical variable $u$.
Consequently,  we have
\begin{align}
  \mathcal{C}_{yy}(\mathfrak{q}) = \frac{ 4 K_4}{z_+} \, \lim_{u \to 0} \frac{a'_y(u|\mathfrak{q})}{a_y(u|\mathfrak{q})} \, .
  \label{eq:25}
\end{align}
The following sections will elucidate the solution of the Einstein-Maxwell equations specified above along with the properties of the temperature-dependent function $\mathcal{C}_{yy}(\mathfrak{q})$.

\subsection{ Master-fields and decoupled equations of motion}
\label{sec:mast-fields-deco}

The linearized Einstein-Maxwell equations in the odd parity sector, eqs.(\ref{eq:51}), can be transformed into a pair of decoupled differential equations for the so-called
master-field $\Phi_\pm$, Ref.\cite{Edalati2010a}\cite{Kodama}:
\begin{align}
    \Phi_\pm'' + \frac{ f' }{ f } \Phi_\pm' - \frac{ Q^2 }{ f }  M_\pm(u|k) \Phi_\pm = 0  \, ,
\label{eq:7}
  \end{align}
  with
\begin{equation}
M_\pm(u|k) := (k^2 -Z^2) + 2u(\pm k -Z) + 4u^2 \, ,
\label{Mfunction}
\end{equation}
from which the fluctuations $a_y$ and ${h_{\, t}^y}$ can be extracted according to
\begin{equation}
  \begin{aligned}
    \Phi_\pm(u | \mathfrak{q}) &= 2 Q^2 [ 2 u - ( Z \pm k) ] a_y - \frac{\mu}{u}\, {h_{\, t}^y}'  \, ,
\label{eq:8}
  \end{aligned}
\end{equation}
Eliminating ${h_{\, t}^y}'$ from (\ref{eq:8}), we obtain that
\begin{align}
  a_y(u|\mathfrak{q}) = \frac{1}{4 Q^2} \frac{1}{k} \left[ \Phi_-(u|\mathfrak{q}) - \Phi_+(u|\mathfrak{q}) \right] \, .
\label{eq:55}
\end{align}
Moreover, as was discussed in \cite{Yin2016}, the solution for ${h_{\, t}^y}$ under the homogeneous boundary condition at $u=0$ vanishes as  ${h_{\, t}^y} \big|_{u \to 0}=O(u^3)$,
and we obtain the relation from (\ref{eq:8}):
\begin{align}
  \big[k + Z \big] \Phi_- + \big[ k - Z \big] \Phi_+ = 0, \quad \text{as} \; u \to 0 \,  .
  \label{eq:39}
\end{align}
The notations in Eqn.(\ref{eq:8})-(\ref{eq:39}) emphasize that $a_y$ and $\Phi_\pm$ are functions of $k$ or $\mathfrak{q}$ as well and the analyticity with respect to
$\mathfrak{q}$ is the main theme of this paper.

Because of the complexity of the Eqn.(\ref{eq:7}), it's impossible to find out explicit solutions for arbitrary momentum $\mathfrak{q}$. Asymptotic solutions can be obtained, however,
for small or large magnitude of $\mathfrak{q}$, and can shed lights on the analyticity.
For a small momentum $q$, the master-field equations turn into
  \begin{align}
    \Phi_\pm'' + \frac{f'}{f} \Phi_\pm' - \frac{1}{f} \bigg[  \frac{f'}{u} + 2 Q^2 u(Z \pm Z) \bigg]  \Phi_\pm = \underbrace{ \frac{Q^2\mathfrak{q}^2}{f}\left(1 \pm \frac{u}{Z} \right) \Phi_\pm + O(\mathfrak{q}^4) }_{ R_\pm(u | \mathfrak{q}^2) } \; ,
 \label{eq:71}
  \end{align}
where the leading order equations
\begin{equation}
  \Phi_\pm'' + \frac{f'}{f} \Phi_\pm' - \frac{1}{f} \bigg[  \frac{f'}{u} + 2 Q^2 u(Z \pm Z) \bigg]  \Phi_\pm = 0
  \end{equation}
are exactly soluble and the subsequent corrections can be figured out perturbatively. For a large momentum $q$, it is convenient to transform the master equation into a Schr\"{o}dinger-like
equation via $\Phi_\pm = \frac{1}{\sqrt{f}}\phi_\pm$, i.e.
\begin{align}
 \phi_\pm'' -  V_\pm(u | k, Q^2) \phi_\pm = 0 \; ,
\label{eq:13}
\end{align}
where the “potential energy”
\begin{align}
  V_\pm(u | k , Q^2) = \frac{Q^2}{f} \left[ (k \pm u)^2 - Z^2 - 6 Zu+ 9u^2 - \frac{4 Q^2 u^4(u-Z)^2}{f} \right]  \,  ,
  \label{eq:57}
\end{align}
is dominated by the first term inside the bracket as $q\to\infty$ and the WKB approximation
 \begin{align}
 (\phi_\pm)_{_\text{WKB}}'' - \frac{Q^2}{f}[k \pm u]^2 (\phi_\pm)_{_\text{WKB}} = 0
\label{eq:45}
\end{align}
becomes handy then. This approximation is particularly useful to locate the Friedel-like singularities of $\mathcal{C}_{yy}(\mathfrak{q})$
for a large imaginary-part of the complex-momentum $\mathfrak{q} \sim k$. The details of the solutions of both cases, small $\mathfrak{q}$ and large $\mathfrak{q}$, will be presented in Sec.~\ref{sec:magn-susc-at}.

\section{ The Analyticity of the Holographic Magnetic Susceptibility}
\label{sec:magn-susc-gaug}

In this section, we shall explore the analyticity of the correlator ${\mathcal C}_{yy}(\mathfrak{q})$ on the complex $\mathfrak{q}$-plane.
It follows from Eqn.(\ref{eq:25}) and (\ref{eq:55}) that the magnetic susceptibility can be singular in two ways: (1) The boundary value of the master field
$\Phi_\pm(0|\mathfrak{q})$ itself is singular. (2) $a_y$ vanishes on the AdS-boundary. The former possibility will be ruled out on the entire physical Riemann sheet (defined below) in $\mathfrak{q}$-plane besides four branch points at zero temperature and on the entire $\mathfrak{q}$-plane at nonzero temperature in the subsection \ref{sec:analyt-magn-susc} below. The latter possibility will be ruled out along the real axis at an arbitrary temperatures in the subsection \ref{sec:absence-nontr-solut}.

\subsection{The analyticity of the solutions of the master field equations}
\label{sec:analyt-magn-susc}

 Considering different singularity structures of the master field equations at zero  and  nonzero temperatures, we treat the two cases separately.

\subsubsection{ Zero temperature case }
\label{sec:zero-temp-case}

At $T=0$, $Q^2=3$, the RN-AdS black hole becomes extremal, and the metric function $f(u)$ in the background solution (\ref{eq:35}) reads
\begin{equation}
  f(u)=f_0(u)\equiv(1-u)^2(1+2u+3u^2) .
\end{equation}
and each master field equation of (\ref{eq:7}) becomes a Fuchs equation with four regular points, $u=1, \frac{1}{3}(-1\pm\sqrt{2}i)$ and $\infty$, which can be transformed into the standard Heun equation. The indices at the horizon ($u=1$) read
\begin{align}
  \triangle_{\pm ,(\pm)} = \frac{ 1 }{ 2 } \left[ -1 (\pm) \sqrt{1 + 2 (k \pm 1)^2} \right] , \quad k \equiv \sqrt{\mathfrak{q}^2+1} \in \mathbb{C} \quad ,
  \label{eq:56}
\end{align}
with $\triangle_{+ ,(\pm)}$ for $\Phi_+$ and $\triangle_{- ,(\pm)}$ for $\Phi_-$ \footnote{ The near horizon geometry at extremality is AdS$_2 \times R^2$ , where $\triangle_{\pm,(\pm)}$ is also the scaling exponent in the IR physics on the boundary theory. However, for the static case, the characteristic of the polarization is manifested in complex momentum space.  }, and produce the asymptotic solutions near the horizon
\begin{equation}
  \Phi_\pm(u|\mathfrak{q})\sim (1-u)^{\triangle_{\pm ,(\pm)}} \, .
  \label{eq:11}
\end{equation}
For a real $k$, the indices $\triangle_{\pm ,(-)}<-1$ give rise to a divergent solutions at the horizon, which in turn generates divergent on-shell actions through the $F^2$ term in the integrand of Eqn.(\ref{eq:2}):
\begin{equation}
  \sqrt{-g}F^2\sim g^{uu}a_y^{\prime 2}\sim (1-u)^{2\triangle_{\pm ,(-)}}  \, .
\end{equation}
Consequently, only the positive indices, $\triangle_{\pm ,(+)}$, should be retained, which give rise to a finite on-shell action.
For an arbitrary complex $k$, we may replace (\ref{eq:56}) with
\begin{align}
  \alpha_{\pm} = \frac{ 1 }{ 2 } \left[ -1 + \sqrt{1 + 2 (k \pm 1)^2} \right] , \quad k \equiv \sqrt{\mathfrak{q}^2+1} \in \mathbb{C} \quad ,
  \label{eq:56_1}
\end{align}
supplemented with the requirement ${\rm Re}\{\alpha_{\pm}\}>-1/2$, i.e.,$\rm{Re}\{ \sqrt{1 + 2 (k \pm 1)^2} \}>0$ for a finite action solution. This defines the physical
Riemann sheet of the square root on the complex $k$-plane, being cut along the lines where $\sqrt{1 + 2 (k \pm 1)^2}$ becomes imaginary, i.e.
\begin{equation}
k=\mp1\pm\frac{i}{\sqrt{2}}\eta  \qquad  \eta\in[1,\infty)
\label{cut}
\end{equation}
originated from the branch points $k=\mp1\pm\frac{i}{\sqrt{2}}$.

Consider $\Phi_+(u|\mathfrak{q})$ first. Introducing a new variable $v = 1- u$, and writing
\begin{align}
  \Phi_+(u|\mathfrak{q}) = C_+ \, v^{\alpha_+} P_+(v|k ) \; ,
  \label{eq:75}
\end{align}
the master-field equation $\Phi_+$ in Eqn.(\ref{eq:7}) is transformed into a Heun-type equation for $P_+(v|k)$:
\begin{align}
  v(a v^2 + b v + c) P_+'' + (\beta v^2 + \gamma z + \delta) P_+' + (r v + s) P_+ = 0,
  \label{eq:76}
\end{align}
with the coefficients given by
\begin{equation}
  \begin{aligned}[rlrlrl]
    a &= 3, &\; b &= -8, &\; c &= 6 ; \\
    \beta &= 6(\alpha_+ + 2), &\; \gamma &= -8(2 \alpha_+ +3), &\; \delta &= 12(\alpha_+ +1), \\
    r &= 3\alpha_+ (\alpha_+ +3), &\; s &= -8 \alpha_+ (\alpha_++2) + 6 (3 + k) \; .
  \end{aligned}
  \label{eq:77}
\end{equation}
This equation can be solved by an infinite series
\begin{align}
P_+(v|k) = \sum_{n=0}^\infty G_n(k) v^n \,  ,
\label{eq:78}
\end{align}
with the recurrence relation among successive coefficients
\begin{align}
  G_1 &= - \frac{s}{\delta} \\
  G_{n+1} &= - \frac{n(n-1)b + n \gamma + s}{(n+1)(nc + \delta)} G_n - \frac{(n-1)(n-2)a + (n-1) \beta + r}{(n+1)(nc + \delta)} G_{n-1}   \quad ,
            \label{eq:79}
\end{align}
where we have set $G_0 =1$. Evidently, the denominators in (\ref{eq:79}) cannot vanish on the physical Riemann sheet of the complex $k$-plane
characterized by ${\rm Re}\alpha_+>-1/2$ and each coefficient of the infinite series is analytic there. Poles of recursion coefficients will show up
in un-physical Riemann sheets, where $\mathrm{Re} \{ \alpha_+\}\in(-\infty,-1/2)$.

On the other hand, the distance from the regular point $v=0$ to the nearby regular points $v_\pm=1-\frac{1}{3}(-1\mp\sqrt{2}i)=\frac{1}{3}(4\pm\sqrt{2}i)$ is greater than one, implying that the AdS-boundary, $v=1$, is inside the convergence circle of the power series (\ref{eq:78}). It follows that the infinite series $P_+(1|k)$ and
its derivative with respect $v$ at the boundary converge uniformly with respect to a finite $k$ and thereby share the same analyticity with their coefficients
$G_n(k)$. To demonstrate this point, the infinite series (\ref{eq:78}) is splitted into the sum of its first $N$
terms, $P_+^{(N)}(v|k)$, and the remainder $R_+^{(N)}(v|k)$, i.e. $P_+(v|k)=P_+^{(N)}(v|k)+R_+^{(N)}(v|k)$ with
\begin{align}
  P_+^{(N)}(v|k) \equiv   \sum_{i = 0}^N G_i(k) v^i    ; \qquad  R_+^{(N)}(v|k) =\sum_{i=1}^\infty G_{N+i}(k) v^{N+i} \quad .
\label{eq:78}
\end{align}
For $N \gg \text{max} \{ 1,|k| \}$, the recursion formula Eqn.~(\ref{eq:79}) for the coefficients of $R_+^{(N)}(v|k)$ becomes approximately
\begin{align}
  G_{N+(i+1)} = \frac{4}{3} G_{N+i} - \frac{1}{2} G_{N+(i-1)} , \quad \text{for} \quad  N \gg 1 \quad,  \label{eq:6_}
\end{align}
that implies asymptotic expression of $G_{N+n}$ from Eqn.~(\ref{eq:79})\footnote{The derivation is left in Appendix \ref{sec:asympt-expr-g_nvk}},
\begin{align}
  G_{N+n}(k) = \frac{\sqrt{2}}{4}\mathrm{i} \bigg[ 3 G_N(k)\big( v_-^{1-n} - v_+^{1-n} \big) + \big(3 G_{N-1}(k) - 8 G_N(k) \big) \big(v_-^{-n} - v_+^{-n} \big)\bigg] \quad ,
  \label{eq:4}
\end{align}
for $n = 1,2,3, \cdots$ in terms of $G_N$ and $G_{N-1}$. Since $|v_\pm^{-1}|<1$, we have
\begin{align}
 |G_{N+n}| \sim |v_\pm^{-1}|^n  \to 0 , \quad \text{as} \quad n \to \infty  \;  . \label{eq:5}
\end{align}
For a finite $k$, say, $|k|<K$, there is always a $k$-independent $N$ such that (\ref{eq:6_}) approximates to a specified accuracy. In addition, we can always
find $k$-independent upper bound of $|G_N|$ and $|G_{N-1}|$ and thereby a $k$-independent upper bound of the remainder for a given $|v|<|v_\pm|$.
Consequently,   we end up with two uniformly convergent series
\begin{equation}
  P_+(1|k)=\sum_{n=0}^\infty G_n(k)  \, ; \qquad   P_+^\prime(1|k)=\sum_{n=1}^\infty n G_n(k)  \,  ,  \label{eq:52}
  \end{equation}
with respect to $k$,  which is thereby analytic on the physical Riemann sheet of the complex $k$-plane.

It follows from (\ref{Mfunction}) and (\ref{eq:56_1}), and the analysis given above that
\begin{equation}
M_-(k)=M_+(-k)  \qquad  \alpha_-(k)=\alpha_+(-k)
\label{reflection}
\end{equation}
and both $P_-(1|k)$ and $P_-^\prime(1|k)$ are also analytic on physical Riemann sheet of the complex $k$-plane.

Now we construct the correlation function $\mathcal{C}_{yy}(\mathfrak{q})$ at zero temperature.
It follows from (\ref{eq:55}) and (\ref{eq:39})  that
  \begin{align}
    \frac{C_+}{C_-} =\frac{(1 + k)P_-(1|k)}{(1 - k)P_+(1|k)}
  \end{align}
  and
  \begin{align}
    a_y(0|k) = \frac{C_-}{6(k - 1)} P_-(1 | k) \,  ,
\end{align}
hence,
\begin{equation}
\begin{aligned}
  \mathcal{C}_{yy}(\mathfrak{q})&=\frac{2K_4}{z_+k}\left\{ (k+1)\left[\alpha_++\frac{P_+^\prime(1|k)}{P_+(1|k)}\right]
                                +(k - 1)\left[\alpha_-+\frac{P_-^\prime(1|k)}{P_-(1|k)}\right]\right\}\\
                               &=\frac{2K_4}{z_+k}\left\{(1+k)\left[\alpha(k)+\frac{P^\prime(1|k)}{P(1|k)}\right]
                                -(1 - k)\left[\alpha(-k)+\frac{P^\prime(1|-k)}{P(1|-k)}\right]\right\}  \;,
\end{aligned}
\label{even}
\end{equation}
where we have removed the subscript "+" of $\alpha_+(k)$ and $P_+(1|k)$, and write $\alpha_-(k)=\alpha(-k)$ and $P_-(1|k)=P(1|-k)$ with the aid of
(\ref{reflection}). As the RHS of (\ref{even}) is an even function of $k$, the transformation $k=\sqrt{\mathfrak{q}^2+1}$ will not
introduce new branch points on the physical Riemann sheet of the complex $\mathfrak{q}$-plane, which remains characterized by the branch cuts (\ref{cut}).
The trajectories of the branch cuts (\ref{cut}) on the complex $k$-plane and the relation $\mathfrak{q}^2= k^2 -1$ implies the following
trajectories of the same set of branch cuts on the complex $\mathfrak{q}$-plane
\begin{equation}
  \left\{
    \begin{aligned}
  (\mathrm{Re}\ \mathfrak{q})^2 - ( \mathrm{Im}\ \mathfrak{q})^2 &= -\frac{\eta^2}{2} \\
  \mathrm{Re}\ \mathfrak{q}\cdot \mathrm{Im}\ \mathfrak{q} &= \pm\frac{\eta}{\sqrt{2}}
\end{aligned}
\right.   \quad .   \label{eq:53}
\end{equation}
or explicitly
 \begin{equation}
  \left\{
    \begin{aligned}
  {\rm Re} \ \mathfrak{q} &= \pm\frac{\eta}{2}\sqrt{\sqrt{1+\frac{8}{\eta^2}}-1} \\
  {\rm Im}\ \mathfrak{q} &= \pm\frac{\eta}{2}\sqrt{\sqrt{1+\frac{8}{\eta^2}}+1}
\end{aligned}
\right.   \quad .  \label{eq:54}
\end{equation}
with $\eta=[1,\infty)$. The end-points of the branch-cuts (the branch points) are given by $\eta=1$, which gives rise to
\begin{align}
  \mathfrak{q}_{_\text{end-point}} = \pm \frac{1}{\sqrt{2}}  \pm \mathrm{i} \; .
  \label{eq:58}
\end{align}

\subsubsection{Nonzero temperature}

The master equation at a nonzero temperature is a Fuchs equation of five regular points with the horizon $u=1$ one of them and none of the others
falls within the section of the real axis $u \in [0,1]$ between the boundary and the horizon. As was shown in Ref.\cite{Yin2016},
both indices at the horizon are zero and a power series in $v=1-u$ can be developed for the finite action solution with all coefficients polynomials in
$\mathfrak{q}$. Unlike the zero temperature case discussed above, the circle of convergence of this series may or may not extend beyond the AdS boundary $v=1$. If not,
the AdS boundary can be reached by a sequence of analytic continuations bridging the power series solution around the horizon with the power series
solutions around ordinary points of the differential equation along the line from $v=1$ to $v=0$. Consequently, it was shown that the solution at the AdS boundary is an analytic function
in any finite domain on the complex $\mathfrak{q}$-plane and the correlation function ${\mathcal C}_{yy}(\mathfrak{q})$ is a meromorphic function.

\subsection{The absence of nontrivial solutions with vanishing $a_y$ at the AdS boundary}
\label{sec:absence-nontr-solut}

According to the definition of the master-fields (\ref{eq:8}), a nontrivial solution of $a_y(u|\mathfrak{q})$ that vanishes at the AdS-boundary, $u=0$, implies that
both master fields vanish at the AdS-boundary except for the case $\mathfrak{q}=0$ ($Z-k=0$)\footnote{At $Z=k$, $\Phi_-$ vanishes
at $u=0$ even $a_y$ takes a nonzero finite value there.} and at least one of them is nontrivial off the boundary(This property is reflected in the explicit construction (\ref{even}) at zero temperature.). To rule out such a possibility,
we start with the modified master-field equations (\ref{eq:13}), and find that the nontrivial solutions $\phi_\pm$ of the modified master-field equations which contribute to the
poles of ${\mathcal C}_{yy}(\mathfrak{q})$ correspond to the solutions $\phi_\pm= \sqrt{f} \,\Phi_\pm$ of (\ref{eq:13}) under the Dirichlet boundary conditions
\begin{align}
  \lim_{u \to 0} \phi_ - &= \lim_{u \to 0} \phi_+ = 0  \, ; \label{eq:9}\\
  \lim_{u \to 1} \phi_ - &= \lim_{u \to 1} \phi_+ = 0  \, , \label{eq:10}
\end{align}
where the AdS-boundary conditions (\ref{eq:9}) follow from eq.(\ref{eq:8}) and the horizon conditions (\ref{eq:10}) result from the regularity requirement of $\Phi_\pm$ there.

The solutions $\phi_\pm(u | k , Q^2)$ of (\ref{eq:13}) together with the boundary conditions (\ref{eq:9}) and (\ref{eq:10}) correspond to
the zero energy eigenstate of the one-dimensional Hamiltonian:
\begin{equation}
  H_\pm=-\frac{\mathrm{d}^2}{\ \mathrm{d}u^2}+V_\pm(u|k,Q^2)  \,  ,
  \label{hamiltonian}
\end{equation}
defined between two infinitely repulsive barriers for $u<0$ and $u>1$. Because of the following two properties of the potential (\ref{eq:57}),
\begin{equation}
  V_+(u|k,Q^2)>V_-(u|k,Q^2)  \, ; \qquad   \frac{\mathrm{d}V_\pm}{\mathrm{d}k}>0 \, , \, \text{for} \; u \in (0,1) \, ,
\end{equation}
an eigenstate of $H_-$ at $k=Z(\mathfrak{q}=0)$ with a negative eigenvalue might be escalated to zero eigenvalue of $H_\pm$ at some $k>Z(\mathfrak{q}\ne 0)$.
If we could rule out the former, we would rule out the zero energy eigenstate in both $H_\pm$ when $\mathfrak{q}\ne 0$.

It is easy to find an explicit solution of the master-field equation for $\Phi_-$ at $k=Z$
\begin{equation}
  \Phi_-^{\prime\prime} + \frac{f^\prime}{f}\Phi_-^\prime - \frac{f^\prime}{u f}\Phi_-=0
\end{equation}
i. e. $\Phi_-=u$. This solution is regular at the horizon and  corresponds to a zero energy eigenstate of $H_-$ at $k=Z$, of Eqn.~(\ref{hamiltonian}), under the Dirichlet conditions,
(\ref{eq:9}) and (\ref{eq:10}), i.e.
\begin{equation}
  \phi_-=u \sqrt{f} \, .
  \label{zero}
\end{equation}
Notice that the wave function (\ref{zero}) of zero eigenvalue does not have zeros for $0<u<1$ at any temperature. According to the theory of the Sturm-Liouville problem defined by the eigenvalue problem $H_-\phi_-=E\phi_-$, the eigenvalue $E$ of any eigenstate orthogonal to (\ref{zero}) has to be positive. An explicit proof of this statement is shown in Appendix \ref{sec:proof-non-negativity}. Consequently, there can not be zero energy eigenstate of $H_\pm$ at $\mathfrak{q}\ne 0$.

It follows from (\ref{eq:8}) that the zero eigenstate of $H_-$ at $k=Z$ itself, however, does not imply a vanishing $a_y(0|\mathfrak{q})$ and thereby does not
imply a singularity of ${\mathcal C}_{yy}(\mathfrak{q})$ at $\mathfrak{q}=0$. The perturbation theory in the next section shows that
${\mathcal C}_{yy}(\mathfrak{q})\sim\mathfrak{q}^2$ as $\mathfrak{q} \to 0$.

The second possibility for the singularity of ${\mathcal C}_{yy}(\mathfrak{q})$ is thereby ruled out for a real $\mathfrak{q}$.

Summarizing this section, we have analytically located the branch cuts of the holographic magnetic susceptibility on the physical sheet of the complex-$\mathfrak{q}$ plane at zero temperature and proved rigorously the absence of poles on the real axis at any temperature. What we have not succeeded is to rule out poles on the physical Riemann sheet away from the real axis at
zero temperature.

\section{Magnetic Susceptibility at a Small Momentum and a Large Momentum}
\label{sec:magn-susc-at}
\subsection{ Small momentum expansion}
\label{sec:magn-susc-with}

\subsubsection{ Exact solutions of the master-fields at $\mathfrak{q}=0$ }
\label{sec:regul-solut-mast}

The ``inhomogeneous'' equations in Eqn.(\ref{eq:71}) facilitate an iterative procedure to find the perturbative solutions for small $\mathfrak{q}$, provided that
their homogeneous parts, $\mathfrak{q}=0$ case,
\begin{align}
  {\Phi_\pm^{(0)}}'' + \frac{f'}{f} {\Phi_\pm^{(0)}}' - \frac{1}{f} \bigg[  \frac{f'}{u} + 2 Q^2 u(Z \pm Z) \bigg]  \Phi_\pm^{(0)} = 0 ,
  \label{eq:17}
\end{align}
are explicitly solvable, which is indeed the case. It is easy to verify the following particular solutions
\begin{equation}
  \begin{aligned}
    {\chi_+}^{(0)} &= Z-u \\
    {\chi_-}^{(0)} &= u  \, ,
  \end{aligned}
  \label{eq:18}
\end{equation}
for $\Phi_+^{(0)}$ and $\Phi_-^{(0)}$, respectively. The leading order solutions of the master field equations that are regular at the horizon read then
\begin{equation}
  \Phi_\pm^{(0)}=a_\pm{\chi_\pm}^{(0)}.
  \label{leading}
\end{equation}
The other linearly-independent particular solutions of (\ref{eq:17}), denoted as $\eta_\pm^{(0)}$, can be obtained from the Wronskians of (\ref{eq:17}), i.e.
\begin{align}
  W[ \chi^{(0)}_\pm, \eta^{(0)}_\pm](u)= \text{const.} \; \exp \left( - \int_0^u \frac{f'(s)}{f(s)} \ \mathrm{d}s \right)  \equiv \frac{1}{f(u)} \; ,
  \label{eq:19}
\end{align}
where the arbitrary multiplicative constant is set as $1$. Solving the first-order differential equations in Eqn.~(\ref{eq:19}), i.e.,
\begin{align}
  \chi_\pm^{(0)} (\eta_\pm^{(0)})' - (\chi_\pm^{(0)})' \, \eta_\pm^{(0)} = \frac{1}{f} \,   ,
\end{align}
we find that
\begin{equation}
  \begin{aligned}
    \eta_+^{(0)} &= (Z-u) \int_0^u \frac{\mathrm{d}\xi}{(Z-\xi)^2 \, f} \\
    \eta_-^{(0)}  &= -1 + u \int_0^u \frac{1}{\xi^2}\left( \frac{1}{f}-1 \right) \mathrm{d}\xi  \, .
  \end{aligned}
  \label{eq:20}
\end{equation}
While the integrations involved in Eqn.(\ref{eq:20}) are all elementary, the explicit forms of $\eta_\pm^{(0)}$ are cumbersome and both of them are singular
at the horizon.

\subsubsection{ Fluctuation $a_y$ and magnetic susceptibility up to $\mathfrak{q}^2$-order  }
\label{sec:fluctuation-a_y-up}

Employing the method of variation of parameters,  we obtain a pair of particular solutions of (\ref{eq:71}):
\begin{align}
  (\Phi_\pm)_{_\text{P.S}} &=  - \frac{Q^2 \mathfrak{q}^2}{Z} \left [ \chi_\pm^{(0)}   \int_0^u  \eta_\pm^{(0)} \chi_\pm^{(0)} ( Z \pm s)  \mathrm{d}s + \eta_\pm^{(0)} \int_u^1\big(  \chi_\pm^{(0)} \big)^2 (Z \pm s) \mathrm{d}s \right] + O(\mathfrak{q}^4) \, ,
                             \label{eq:21}
\end{align}
which is regular at the horizon and serves the next order correction to (\ref{leading}). Combining (\ref{leading}) and (\ref{eq:21}), we find the
solutions of the master-field equations~(\ref{eq:71}) in small momentum approximation:
\begin{align}
  \Phi_\pm = a_\pm \, \chi_\pm^{(1)} + O(\mathfrak{q}^4) \, ,  \label{eq:37}
\end{align}
where
\begin{equation}
  \label{eq:30}
  \begin{aligned}
    \chi_+^{(1)} &= (Z-u) - \frac{Q^2\mathfrak{q}^2 }{Z} \bigg[  (Z-u) \int_0^u \eta_+^{(0)}(Z^2 - s^2)\,  \mathrm{d}s  + \eta_+^{(0)} \int_u^1 (Z + s) (Z - s)^2 \, \mathrm{d}s \bigg]  \\
    \chi_-^{(1)} &= u - \frac{Q^2 \mathfrak{q}^2}{Z} \left[ u \int_0^1 \eta_- ^{(0)}(Z -s) s \, \mathrm{d}s + \eta_-^{(0)} \int_u^1 (Z - s) s^2 \,  \mathrm{d}s \right] \,  .
  \end{aligned}
\end{equation}
The two coefficients in Eqn.~(\ref{eq:37}) are not arbitrary and are constrained by the behaviour of the perturbed metric fields $h_{\, t}^y$ as $u\to 0$, which implies (\ref{eq:39}).
Substituting (\ref{eq:37}) and (\ref{eq:30}) into (\ref{eq:39}), we obtain the ratio of the two coefficients
\begin{align}
  \frac{ a_-  }{ a_+ }  = -1 \, .
  \label{eq:40}
\end{align}
Following eqs. (\ref{eq:37}) and (\ref{eq:55}),  the fluctuation $a_y(u|\mathfrak{q})$ and its derivative w.r.t $u, a_y'(u|\mathfrak{q})$,  take the form:
\begin{equation}
  \begin{aligned}
    a_y (u|\mathfrak{q})&= \frac{a_-}{ 2 Q^2} \frac{1}{k} \left[   \left( \chi_-^{(1)} + \chi_+^{(1)} \right) + O(\mathfrak{q}^4)   \right]\\
    &= \frac{a_-}{ 2 Q^2} \frac{1}{k}  \bigg\{
    \left(    u - \frac{Q^2 \mathfrak{q}^2}{Z} \left[ u \int_0^u \eta_-^{(0)} (Z - s) s \, \mathrm{d}s + \eta_-^{(0)} \int_u^1 (Z- s) s^2  \, \mathrm{d}s \right]\right)  \\
  & \qquad \qquad + \bigg( (Z-u) - \frac{Q^2 \mathfrak{q}^2}{Z} \bigg[  (Z - u) \int_0^u \eta_+^{(0)} (Z^2 - s^2)  \, \mathrm{d}s    \\
  &\hspace{5cm} + \eta_+^{(0)} \int_u^1 (Z + s)(Z - s)^2 \, \mathrm{d}s \bigg] \bigg) + O(\mathfrak{q}^4)
  \bigg\}
\end{aligned}
\label{eq:41}
\end{equation}
and
\begin{equation}
  \begin{aligned}
    a_y'(u|\mathfrak{q}) &= \frac{a_-}{ 2 Q^2} \frac{1}{k}  \bigg\{
    \bigg(  1 - \frac{Q^2 \mathfrak{q}^2}{Z} \bigg[   \int_0^u \eta_-^{(0)} (Z - s) s \, \mathrm{d}s + \eta_-^{(0)} (Z-u) u^2   \\
    & \hspace{2cm} + \left(\eta_-^{(0)}\right)' \int_u^1 (Z-s)s^2 \, \mathrm{d}s - \eta_-^{(0)} (Z-u)u^2   \bigg]\bigg)  \\
    & \hspace{0.8cm} + \bigg( -1 - \frac{Q^2 \mathfrak{q}^2}{Z}  \bigg[  - \int_0^u \eta_+^{(0)} (Z^2-s^2) \, \mathrm{d}s     + \eta_+^{(0)}(Z-u)(Z^2-u^2)  \\
    & \hspace{0.8cm}  + \left( \eta_+^{(0)}\right)' \int_u^1 (Z+s)(Z-s)^2 \,  \mathrm{d}s - \eta_+^{(0)} (Z+u)(Z-u)^2
    \bigg]\bigg) + O(\mathfrak{q}^4)
    \bigg\} \, .
  \end{aligned}
  \label{eq:42}
\end{equation}
The overall constant $a_-$ drops in the correlation function $\mathcal{C}_{yy}$ in accordance with Eqn.(\ref{eq:25}) and we obtain that
\begin{equation}
  \begin{aligned}
    \mathcal{C}_{yy}(\mathfrak{q})  &=  \frac{ 4 K_4 }{ z_+ \, Z } \frac{4 Z^2 + [ -4 Z + (Z^2(Q^2 +3) -1)] \mathfrak{q}^2  +  O(\mathfrak{q}^4)}{ \mathfrak{q}^2 + 4 Z^2 +  O(\mathfrak{q}^4) } \\
    & =  \frac{ K_4 }{ z_+ \mu^2  }  \frac{ Z^2(Q^2+3) -1 }{ Z^3 } \, q^2  + O(q^4) \, ,
  \end{aligned}
  \label{eq:43}
\end{equation}
where $q$ is the unscaled momentum. The dimension of the holographic polarization tensor is $[\mathcal{C}_{yy}]=[q^2/\mu]=1$, as expected. Following Eqn.(\ref{eq:32}), the magnetic susceptibility
at zero momentum reads
\begin{align}
  \upchi(q|T) =  \frac{ K_4 }{ z_+ \mu^2  }\frac{Z^2(Q^2 +3)-1}{Q^2 Z^3} \, ,
  \label{eq:12}
\end{align}
which becomes
\begin{align}
  \upchi(q|0) = \frac{5}{3}  \frac{ K_4 }{ z_+ \mu^2  } \, .
  \label{eq:44}
\end{align}
at zero temperature.

\subsection{ WKB approximation at a large momentum }
\label{sec:fluct-a_y-magn}

The region far away from the real momentum-axis can be explored by the WKB-approximation of the modified master-fields $\phi_\pm = \sqrt{f} \Phi_\pm$,  and the fluctuation $a_y$ in 
the WKB-approximation can be obtained from the solutions of the Schr\"{o}dinger-like
equations~(\ref{eq:45}) via the relation (\ref{eq:55}). The nonzero temperature case has been worked out in \cite{Yin2016} and we include the key steps in Appendix A
for self-containedness. There we also derived the asymptotic form of the magnetic susceptibility which was missing in \cite{Yin2016}. In what follows, we shall focus
on the zero temperature case.
Unlike the non-extremal blackhole, the validity of the WKB-approximation extends all the way from the boundary to the horizon because the condition of the approximation,
$|V_\pm^\prime| \ll|V_\pm|^{3/2}$ \cite{weinberg} holds for $0\le u \le 1$.

The general WKB solutions of (\ref{eq:45}) at $T=0$ read
  \begin{equation}
    (\phi_\pm)_{_\text{WKB}} \propto f_0^{1/4}\left[C_\pm\exp \left( \sqrt{3} \int_0^u \frac{k\pm v}{\sqrt{f_0}}\ \mathrm{d}v \right)
      +D_\pm\exp\left( -\sqrt{3} \int_0^u \frac{k\pm v}{\sqrt{f_0}}\ \mathrm{d}v \right)\right]
\label{eq:15}
  \end{equation}
  with $f_0=(1-u)^2(1+2u+3u^2)$. The integrals in the exponents can be carried out explicitly, i.e.
  \begin{align}
    \int_0^u \frac{k+v}{\sqrt{f_0}} \ \mathrm{d}v = (k+1)A(u)-B(u)  \\
    \int_0^u \frac{k-v}{\sqrt{f_0}} \ \mathrm{d}v = (k-1)A(u)+B(u)
  \end{align}
where
  \begin{equation}
    A(u)=\frac{1}{\sqrt{6}}\ln\frac{2+4u+\sqrt{6(1+2u+3u^2)}}{(2+ \sqrt{6})(1-u)}
  \end{equation}
  and
  \begin{equation}
    B(u)=\frac{1}{\sqrt{3}}\ln\frac{1+3u+ \sqrt{3(1+2u+3u^2)}}{1 + \sqrt{3}}
  \end{equation}
Because of the divergence of $A(u)$ as $u\to 1$, one of the terms inside the bracket of Eqn.(\ref{eq:15}) blows up at the horizon
  for $\text{Re}\, k \neq\pm 1$ and has to be dropped for a finite on-shell action. We have
  \begin{equation}
    (\phi_+)_{_\text{WKB}}  \propto f_0^{1/4}
    \left\{
      \begin{aligned}
        &\exp\left\{ -\sqrt{3} [(k+1)A(u)+B(u)] \right\} \, , \; \text{for} \; \text{Re}\,  k >-1 \, ; \\
          &\exp\left\{ \sqrt{3} [(k+1)A(u)+B(u)] \right\}  \, , \; \text{for} \ \text{Re}\, k <-1 \;,
          \end{aligned}
        \right.
      \end{equation}
  and
  \begin{equation}
    (\phi_-)_{_\text{WKB}}  \propto f_0^{1/4}
    \left\{
      \begin{aligned}
        &\exp\left\{ -\sqrt{3} [(k-1)A(u)+B(u)] \right\} \, , \; \text{for} \; \text{Re}\,  k >1 \, ; \\
          &\exp\left\{ \sqrt{3} [(k-1)A(u)+B(u)] \right\}  \, , \; \text{for} \;\text{Re}\, k <1 \;,
          \end{aligned}
        \right.
      \end{equation}
As expected, the discontinuity of the master-field $\phi_-$ at ${\rm Re}k=1$ and the discontinuity of the master field $\phi_+$ at ${\rm Re}k=-1$ match the asymptotic trajectories of the
branch-cuts (\ref{cut}), which correspond to the condensation of the poles discussed in \cite{Yin2016} as $T\to 0$  .

\section{ Discussion and Conclusion }
\label{sec:disc-concl-}

In this work, we explored the analyticity of the static transverse component of the holographic polarization tensor, $\mathcal{C}_{yy}(\mathfrak{q})$, in 2+1 dimensions with respect to the
complex spatial momentum. The dimensionless  magnetic susceptibility is given by $\mathcal{C}_{yy}(\mathfrak{q}) / q^2$. The zero temperature features of the static holographic susceptibility are not determined by the near-horizon IR data, but by the analyticity in the complex momentum plan.
We provided a rigorous proof that $\mathcal{C}_{yy}(\mathfrak{q})$ is analytic in the neighbourhood of a real $\mathfrak{q}$ even at zero temperature. In addition, we located analytically four branch cuts on the complex $\mathfrak{q}$-plane at zero temperature, which terminated at the branch
points $\pm 1/\sqrt{2} \pm \mathrm{i}$, staying away from the real and imaginary axes. We also worked out the asymptotic form of $\mathcal{C}_{yy}(\mathfrak{q})$ for small $\mathfrak{q}$ and large $\mathfrak{q}$. The momentum analyticity of the transverse holographic polarization appears similar to that of the longitudinal one, as was demonstrated by the numerical solution \cite{Blake2015c} and WKB approximation \cite{Yin2017} of the Einstein-Maxwell equations in the sector of even parity \footnote{
    The  distinction between the real part of the branch cuts location in strong coupling and the location of the discontinuity $2\mu$ in weak coupling is observed in other studies \cite{Blake2015c} and \cite{Henriksson2017}. Throughout this paper, we follow the convention in our previous works \cite{Yin2016} and \cite{Yin2017} by scaling the $U(1)$ gauge potential in eq.(\ref{eq:2}) such that $\frac{K_4}{G_4}=L^2=1$. An arbitrary ratio $\frac{K_4}{G_4}\equiv\eta^2$ amounts to the transformations $a_t\to\eta a_t$ and $\mu\to\eta\mu$ in the Einstein-Maxwell equations, and thereby the transform of the asymptotic branch cuts from $k\simeq\mu$ to $k\simeq\mu\eta$. The notation in \cite{Blake2015c} corresponds to $\eta=\frac{1}{2}$ which gives rise to the asymptotic cuts at $q\simeq\pm 1/2 k_F$, while Ref.\cite{Henriksson2017} that starts with ABJM theory gives rise to ``$q \simeq \pm 1k_F$'' result, as oppose to the result in weak-coupled system.}. What we have not achieved is to rule out analytically the poles of $\mathcal{C}_{yy}(\mathfrak{q}) / q^2$ away from the real axis and the branch cuts at zero temperature.

As the holographic polarization tensor may reflect certain strong-coupling properties, it is instructive to compare
our results with the polarization tensor in weak coupling at zero temperature to find their difference. The transverse component of the static polarization tensor in the massless spinor QED at one loop order reads
\begin{equation}
  \sigma^{\text{{tr}}}(q)  =
  \left\{
    \begin{aligned}
      & 0,\  |q| < 2 \mu \, ; \\
      & \frac{q}{16}+\frac{\mu}{4\pi}\sqrt{1-\frac{4\mu^2}{q^2}}-\frac{q}{8\pi}\sin^{-1}\frac{2\mu}{q}, |q| > 2\mu \;,
    \end{aligned}
  \right.
  \label{weak}
\end{equation}
and $\sigma^{\text{tr}}(q)$ is the weak-coupling counterpart of the holographic $\mathcal{C}_{yy}(\mathfrak{q})$.
\begin{figure}[!htbp]
  \centering{
    \subfigure[Weak-Coupling]{
      \begin{minipage}[t]{0.4\linewidth}
        \includegraphics[width=2.5in]{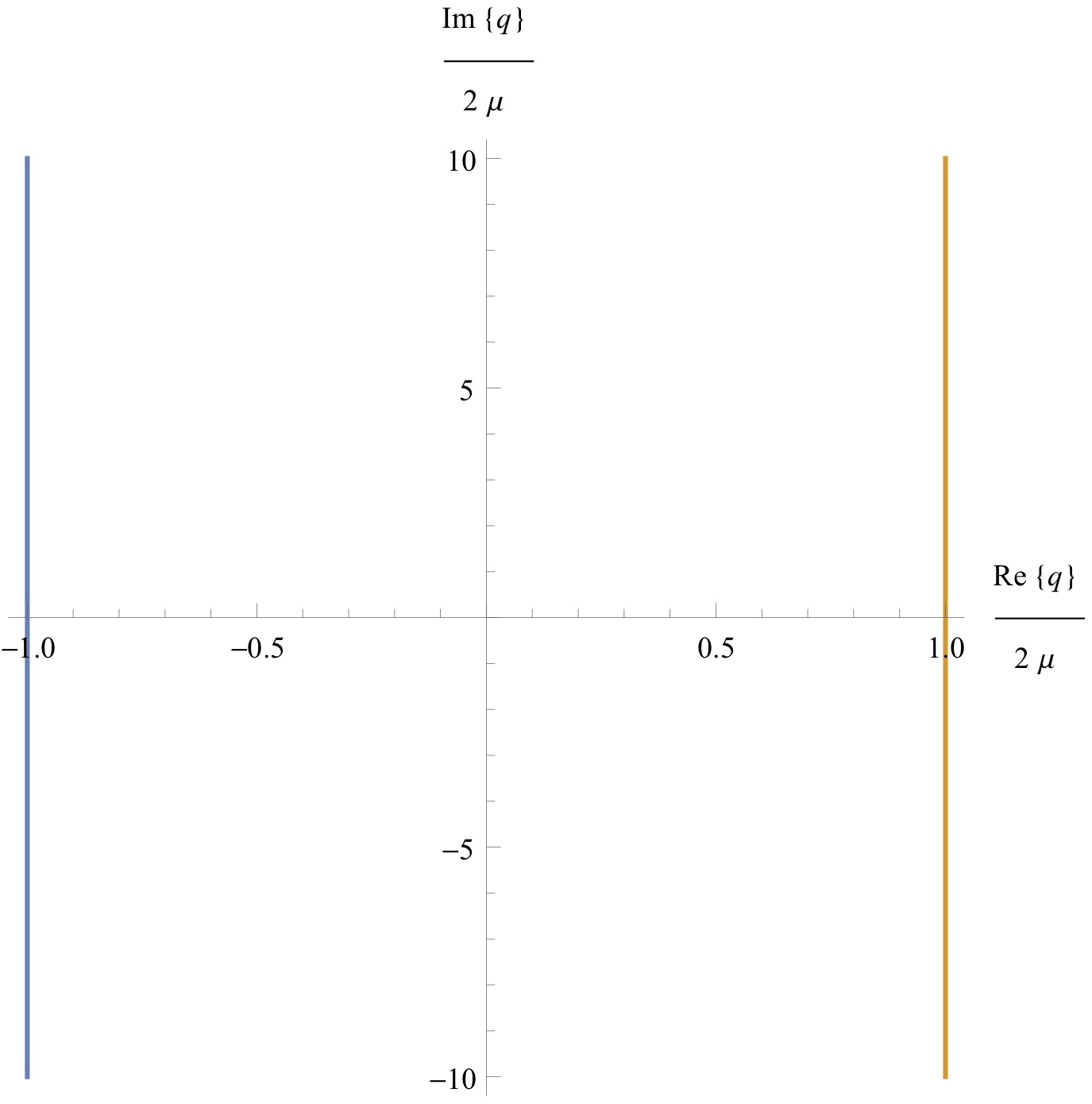}
      \end{minipage}%
    }
    \subfigure[Strong-Coupling]{
      \begin{minipage}[t]{0.4\linewidth}
        \includegraphics[width=2.5in]{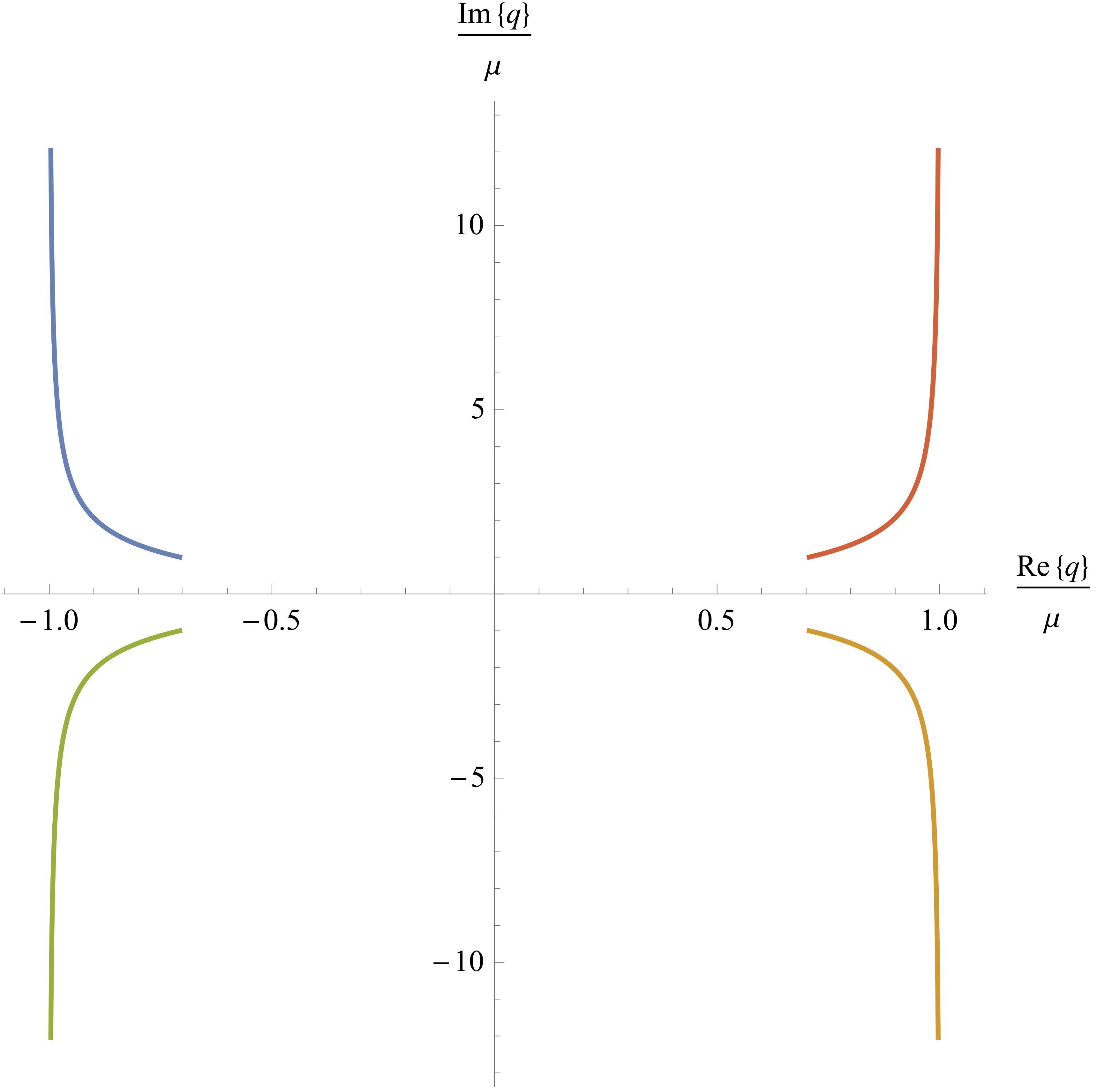}
      \end{minipage}%
    }}
  \caption{Momentum analyticity of static transverse polarization in weak/strongly coupling cases. The lines or curves represent the branch-cuts resulting from the singularities condensation. In weak-coupling (a), the H-type structure of branch-cuts based on Eqn.(\ref{weak}) shows that all branch-points
    condensate on the two kinks $\pm 2\mu$ and indicate a power-low decay mode in a large distance at $T=0 K$; while in strong-coupling (b) depicted by Eqn.(\ref{eq:53}) and (\ref{eq:54}), all poles condensate at $\pm 1/\sqrt{2} \pm \mathrm{i}$ so as the branch-cuts keep a particular separation from the real-axis. }
  \label{fig: comparison}
\end{figure}

The comparison between the momentum analyticity of $\mathcal{C}_{yy}(\mathfrak{q})$
in strong coupling and that of $\sigma^{\text{tr}}(q)$ in weak-coupling at $T=0$ is depicted in Fig.\ref{fig: comparison}.

The left panel in Fig.~\ref{fig: comparison} shows discontinuities in the derivative of $\sigma^{\text{tr}}(q)$
at $q=\pm 2\mu$ because of the condensation of the branch-points (\ref{eq:1}), which form  a pair of branch-cuts crossing the real momentum axis and cause the Friedel oscillation in coordinate space whose
amplitude decays with distance according to a power law. In contrast, the right panel in Fig.~\ref{fig: comparison} for $\mathcal{C}_{yy}(\mathfrak{q})$ shows that
the real axis is free from singularities and is spared by the bending branch-cuts.
The nonzero real parts of the branch cut locations give rise to oscillatory behavior in coordinate space, while the nonzero imaginary parts of them imply exponential decay of the amplitude of the oscillation at large distance in coordinate space even at zero temperature, instead of the power-law decay in weak-coupling case.

To elaborate the observation perspectives of the (2+1)-dimensional polarization tensor in a (3+1)-dimensional environment, we consider the photon propagator
$D_{\mu\nu}(\vec q|z,z')$ in the presence of a sheet of medium located at $z=0$, whose electromagnetic property is described by the holographic polarization tensor.
Here we use the momentum representation in $(x,y)-$ directions and the coordinate representation in $z$ because of the lack of translation invariance in that direction.
The vector potential at a point in the medium in response to a current element located at another point in the medium corresponds to the Fourier transformation
of its transverse component, i.e.
\begin{equation}
  D_{ij}(\vec x)=\int\frac{\ \mathrm{d}^2\vec{q}}{(2\pi)^2}e^{\mathrm{i} \vec{q}\cdot\vec{x}} \ \mathcal{D}_{ij}(\vec{q}|0,0) \; .
\end{equation}
To the first order in the (3+1)-dimensional electromagnetic coupling $e^2$, we have
\begin{align}
  \mathcal{D}_{ij}( \vec{q}|0,0) = \mathcal{D}_{ij}^0(\vec{q}) +  e^2\mathcal{D}_{i a}^0(\vec{q}) \left[ \sigma^{\text{tr.}}(\vec{q})\left(\delta_{ab}-\frac{q_aq_b}{|\vec{q}|^2|}\right) \right]  \mathcal{D}_{b j}^0(\vec{q})  \;
,
  \label{eq:62}
\end{align}
where the first term $\mathcal{D}_{ij}^0(\vec{q})$ is the static transverse component of the (3+1)-dimensional free propagator and the magnetization comes from the polarization in the second term that is our main consideration.
Also, in the static case, $\mathcal{D}_{ij}(q|0,0)$ is always contracted with the Fourier component of a stationary electric current and the factor $q_b$ in (\ref{eq:62}) does not contribute because of the current conservation. Hence,  effectively,
$\mathcal{D}_{ij}(\vec q|0,0)=D(q)\delta_{ij}$ with a scalar form factor
\begin{align}
  D(q) = \frac{1}{2 q} + \frac{e^2}{4 q^2} \sigma^{\rm tr.}(q) \; ,
  \label{eq:65}
\end{align}
where the first term on RHS comes from the free propagator and the second term reflects the polarization of the medium with $e$ the electric charge in 3+1
dimensions, $\sigma^{\text{tr.}}(q)$ is the 2D polarization, proportional to $\mathcal{C}_{yy}(q)$ for the holographic polarization in our case.
Assuming that $\mathcal{C}_{yy}(q)$ has no poles on the entire physical Riemann sheet (not just around the real axis) and  only focusing on the second term of (\ref{eq:62}), Its Fourier transform takes the asymptotic form at large $|\vec{x}|$, i.e.
\begin{equation}
\begin{aligned}
   \int \frac{\mathrm{d}^2 q}{4 \pi^2} \ e^{\mathrm{i} \vec{q} \cdot \vec{x}}\left[\frac{1}{4 q^2} \sigma^{\rm tr.}(q)\right]
\sim \frac{1}{|\vec x|^2}e^{-\mu|\vec x|}\cos\left(\frac{\mu}{\sqrt{2}}|\vec x|+\phi\right)  \, ,
\end{aligned}
\label{eq:59}
 \end{equation}
 where the integral in (\ref{eq:59}) is calculated via a contour integration, going along the branch-cuts in the upper half plane of Fig.~\ref{fig: comparison}(b) and $\phi$ is a phase
constant. The details behind (\ref{eq:62}) and (\ref{eq:59}) can be found in Appendix~\ref{sec:two-dimens-curr}. The exponential factor on the right hand side of (\ref{eq:59}) is explicitly in contrast to the case in a weakly-coupled field theory,
such type of Friedel-like oscillation with faster than power-law decay behavior is observed in the density-density correlation in other holographic strongly-coupled systems\cite{Blake2015c,Henriksson2017,Doucot2017} and in the zero fermionic flavor limit: $N_f \to 0$ \cite{Saterskog2017}.

The Friedel-like oscillation caused by the transverse component of the polarization tensor is responsible for the RKKY effect \cite{RKKY}, where the local magnetic field acting on a
nuclear magnetic moment is generated by other nuclear magnetic moments and is polarized by the Fermi sea of electrons. Therefore, the effect discussed in this work may
find its application in the RKKY effect in some 2D metals, whose low-lying excitations are Dirac like, such as a doped graphene.

On the other hand, one may associate the absence of the cuts crossing the real momentum axis
to the bosonic degrees of freedom which may dominate in the boundary field theory. Even in weak coupling, the singularities for scalar QED at one-loop order at \cite{Yin2017}
\begin{equation}
\quad q = \pm 2 \sqrt{(\mu + \mathrm{i} 2 \pi n T)^2 - m^2}  \, ,
\end{equation}
with $m$ the mass of the charged bosons will not condense towards the
real axis for $|\mu|<m$, which is required for the positivity of the quadratic action underlying the perturbation theory in the absence of a Bose condensate, while a singularity
as $|\mu|\to m$ will show up in thermodynamic functions. But in the holographic model considered in this work, the chemical potential appears unconstrained.
Therefore it is likely that the analyticity of the static polarization tensor around the real momentum axis reflects a generic feature of the strongly coupling of a fermionic system
if the gauge/gravity duality holds and it would be interesting to observe the exponentially decayed oscillation at zero temperature in some strongly correlated
electronic systems.

At the moment, we are unable to generalize the analytical works presented above to the case of the longitudinal component of the polarization tensor because of the
technical complexity and hope to report our progress along this line in near future.


\section*{Acknowledgement}

We are grateful to the anonymous referee for his (her) criticism and suggestions which motivated us to enlarge the scope of this work.
L.Yin indebted to Prof. T.K.Lee for helpful discussions. This work is partly supported by the Ministry of Science and Technology of China (MSTC) under the 973 Project No. 2015CB856904(4). L.Yin is supported by NSFC under Grants Nos.11747050 and D-F. Hou \& H.C. Ren  are supported by the NSFC under Grants Nos. 11735007, 11890711.

\appendix

\section{The WKB Approximation at Nonzero Temperature}
\label{sec:odd-mode-wkb}

The pair of equations (\ref{eq:51}) can be decoupled further through the master fields introduced in the subsection \ref{sec:mast-fields-deco} and the WKB solutions, obtained in Ref.\cite{Yin2016},
\begin{align}
  (\phi_\pm)_{_\text{WKB}} = \frac{f^{1/4}}{Q^{1/2}} D_\pm \left[ \mathrm{i} \exp\left( - Q \int_u^1 \frac{k \pm v}{\sqrt{f} } \ \mathrm{d}v \right) + \exp \left( Q \int_u^1 \frac{k \pm v}{\sqrt{f}} \ \mathrm{d}v \right)  \right]  \, ,
  \label{eq:46}
\end{align}
follow together with their derivatives w.r.t $u$:
\begin{align}
  \left(\phi'_\pm \right)_{_\text{WKB}} = f^{1/4} Q^{1/2} (k \pm u) D_\pm \left[ \mathrm{i} \exp\left\{- Q \int_u^1 \frac{k \pm v}{\sqrt{f} } \ \mathrm{d}v \right\} - \exp \left\{ + Q\int_u^1 \frac{k \pm v}{\sqrt{f}} \ \mathrm{d}v \right\} \right] \, ,
\end{align}
where the ratio of the constants $D_\pm$ are fixed by Eqn.(\ref{eq:39}), i. e.
\begin{align}
  \frac{D_+}{D_-} = - \frac{  \mathrm{i} \exp\left( - Q \int_u^1 \frac{k - v}{\sqrt{f} } \ \mathrm{d}v \right) + \exp \left( Q \int_u^1 \frac{k - v}{\sqrt{f}} \ \mathrm{d}v \right)  }{  \mathrm{i} \exp\left( - Q \int_u^1 \frac{k + v}{\sqrt{f} } \ \mathrm{d}v \right) + \exp \left( Q \int_u^1 \frac{k + v}{\sqrt{f}} \ \mathrm{d}v \right)  } \, .
  \label{eq:29}
\end{align}
We have then the boundary values of $\left(\phi_\pm \right)_{_\text{WKB}}$ and $\left(\phi'_\pm \right)_{_\text{WKB}}$, i.e.
\begin{align}
  \lim_{ u \to 0}   \left(\phi_\pm \right)_{_\text{WKB}} &= Q^{-1/2} \, D_\pm \left[ \mathrm{i} e^{-Q [k L_1 \pm L_2]} + e^{+Q[k L_1 \pm L_2] } \right] \label{eq:27} \\
    \lim_{ u \to 0}   \left(\phi'_\pm \right)_{_\text{WKB}}  &= Q^{1/2} k \, D_\pm  \left[ \mathrm{i} e^{-Q [k L_1 \pm L_2]} - e^{+Q[k L_1 \pm L_2] } \right] \, .
\label{eq:28}
\end{align}
Substituting  (\ref{eq:27}) and (\ref{eq:28}) into the expressions of $a_y$, (\ref{eq:55}), and $\mathcal{C}_{yy}$, (\ref{eq:25}), with the ratio (\ref{eq:29}),
the following WKB approximation of
$\mathcal{C}_{yy} (\mathfrak{q})$ is obtained after some algebra:
\begin{align}
  \mathcal{C}_{yy} (\mathfrak{q})\bigg|_{_\text{WKB}} = \frac{4 K_4 Q}{z_+}  k   \, \frac{\cosh[ 2QL_1 k ]}{\sinh[2 Q  L_1 k] + \mathrm{i} \cosh[2Q L_2]} ,
  \label{eq:50}
\end{align}
with $L_1$ and $L_2$ the two temperature-dependent elliptic integrals:
\begin{align}
  L_1 := \int_0^1 \frac{\ \mathrm{d}u}{\sqrt{1-(1+Q^2)u^3+Q^2u^4}} \quad ,\quad
  L_2 := \int_0^1 \frac{1-u \ \mathrm{d}u}{\sqrt{1-(1+Q^2)u^3+Q^2u^4}} \, ,
  \label{eq:47}
\end{align}
As $T \to 0, \; L_1\to 0, L_2 \to 0$ and $L_2/L_1 \to 0$. Obviously the mass dimension of $[\mathcal{C}_{yy}] = [z_+^{-1}]=1$ as expected and the roots of the denominator in Eqn.(\ref{eq:50})
contribute to asymptotic poles of the transverse polarization. It follows from Eqn.(\ref{eq:32}) that the magnetic susceptibility
\begin{align}
  \upchi(q|T)\bigg|_{_\text{WKB}} = \frac{4 K_4 Q}{ \mu^2 z_+ k}   \, \frac{\cosh[ 2QL_1 k ]}{\sinh[2 Q  L_1 k] + \mathrm{i} \cosh[2Q L_2]} ,
  \label{eq:14}
\end{align}
which is of mass dimension $[\upchi]=[(\mu^2 z_+)^{-1}] = -1$  with $\mathfrak{q} = q/\mu \sim k$ for $q \gg \mu$.

\section{Proof of the non-negativity of the eigenvalues }
\label{sec:proof-non-negativity}

The "Hamiltonian operator" in the subsection \ref{sec:absence-nontr-solut} corresponding to the master field $\Phi_-$ at $k=Z$ reads
\begin{align}
  H_- := - \frac{\mathrm{d}^2}{\mathrm{d}u^2} + V_-(u)
\end{align}
with $V_-(u)\equiv V_-(u|k=Z,Q^2)$ and defines the eigenvalue problem
\begin{align}
  H_-  \phi = E \phi
  \label{eq:16}
\end{align}
It can be shown that the operator $H_-$ is hermitian, subject to the boundary conditions
\begin{align}
 \phi(0) = 0 , \quad \text{and} \quad \phi(u) \sim 1 - u \quad \text{as} \; u \to 1^- \, .
\end{align}
We have found an exact zero mode of this eigenvalue problem at $k=Z$ : $\phi_0 = u \sqrt{f} >0$.
\textit{If there is another mode $\phi(u)$ that is orthogonal to $\phi_0$, the associated eigenvalue $E >0$ }. To prove this statement,
we note that the orthogonality between $\phi(u)$ and $\phi_0(u)$ requires that $\phi(u)$ switch its sign somewhere in the interval $0<u<1$. Let us denote by $\zeta$

the first sign switching zero-point away from $u=0$, and, without loss of generality,
assume $\phi >0$ for $0 <u <\zeta$. Obviously
\begin{align}
  \frac{\mathrm{d}\phi}{\mathrm{d}u}\bigg|_{u = \zeta} <0  \, .
\end{align}
It follows from the eigenvalue equations, $H_- \phi_0 = 0$ and $H_- \phi = E\phi$ that
\begin{align}
  -E \int_0^\zeta \ \mathrm{d}u \, \phi_0 \phi = \left( \phi_0 \frac{\mathrm{d}\phi}{\mathrm{d}u} - \frac{\mathrm{d}\phi_0}{\mathrm{d}u} \phi \right) \bigg|_0^\zeta = \phi_0(\zeta) \frac{\mathrm{d}\phi}{\mathrm{d}u}\bigg|_{u=\zeta} <0  \, .
\end{align}
Then the positivity of the integral on LHS, $\int_0^\zeta \phi_0 \phi \, \ \mathrm{d}u >0$, implies that $E>0$. The proof is completed.

\section{Asymptotic Expression of $G_n(k)$ from Generating Function}
\label{sec:asympt-expr-g_nvk}

The remainder $R_+^{(N)}(v|k)$ defined in (\ref{eq:78}) serves the generating function of the coefficients $\{G_n(k)\}$ when its index is large. It follows from the asymptotic recurrence formula
\begin{align}
  G_{N+(i+1)} = \frac{4}{3} G_{N+i} - \frac{1}{2} G_{N+(i-1)} , \quad \text{for} \quad  N \gg 1 \quad,  \label{eq:6}
\end{align}
that
\begin{align}
  R_+^{(N)} = P_+(v|k) - P_+^{(N)}(v|k) = \frac{4}{3} v \big( R_+^{(N)} + G_N v^N \big) - \frac{1}{2}\big( v^2R_+^{(N)} + G_N v^N + G_{N-1} v^{N-1} \big)  \quad ,
  \label{eq:22}
\end{align}
Solving Eqn.~(\ref{eq:22}) for $R_+^{(N)}$, we obtain that:
\begin{align}
  R_+^{(N)}(v|k) = \frac{\sqrt{2}}{4}\mathrm{i} \left( \frac{1}{v - v_-} - \frac{1}{v - v_+} \right) v^N \big[ 3  G_N v^2 + (3 G_{N-1} - 8 G_N) v \big] \quad , \label{eq:23}
\end{align}
where $v_\pm = \frac{1}{3}[4 \pm \mathrm{i} \sqrt{2}]$. The function $R_+^{(N)}$ can be safely expanded according to the power of $\frac{v}{v_\pm}$ for
$|\frac{v}{v_\pm}| <1$, i. e.

\begin{align}
  \tilde{P}_+ &= \frac{\sqrt{2}}{4}\mathrm{i} v^N \bigg[ 3 G_N \sum\limits_{i=2}^{\infty} \left(\frac{1}{v_-^{i-1}}  - \frac{1}{v_+^{i-1}}\right) v^i + (3 G_{N-1} - 8 G_N) \sum\limits_{i=1}^{\infty} \left( \frac{1}{v_-^i} - \frac{1}{v_+^i} \right) v^i  \bigg]  \notag \\
              &=  \frac{\sqrt{2}}{4}\mathrm{i} v^N \bigg\{ (3 G_N - 8 G_N) \left(  \frac{1}{x_-} - \frac{1}{x_+}  \right) v \\
              & \qquad + \sum\limits_{i=2}^{\infty}  \left[  3 G_N \left(  \frac{1}{v_i^{i-1}} - \frac{1}{v_+^{i-1}}\right) + (3 G_{N-1} - 8G_N) \left( \frac{1}{v_-^i} - \frac{1}{v_+^i} \right) \right]  \cdot v^i  \bigg\} \, .
                \label{eq:24}
\end{align}
from which the asymptotic expression of $G_n$ for large indexes $n \to \infty$ can be extracted as
\begin{align}
  G_{N+n}(k) &= \frac{\sqrt{2}}{4}\mathrm{i} \bigg[ 3 G_N(k)\big( v_-^{1-n} - v_+^{1-n} \big) + \big(3 G_{N-1}(k) - 8 G_N(k) \big) \big(v_-^{-n} - v_+^{-n} \big)\bigg]   \label{eq:26} \\
             &= \frac{\sqrt{2}}{4}\mathrm{i} \bigg[ \big( 3 G_N(k) v_- + 3 G_{N-1}(k) - 8 G_N(k) \big) \cdot v_-^{-n}  \notag \\
  & \hspace{4cm}   - \big( 3 G_N(k) v_+ + 3 G_{N-1}(k) - 8 G_N(k) \big) \cdot v_+^{-n} \bigg]  \; ,  \label{eq:36}
\end{align}
Eqn.~(\ref{eq:4}) follows then.


\section{ Observing 2D polarization tensor in 3D environment }
\label{sec:two-dimens-curr}

In the presence of a homogeneous medium in $x-y$ plane, the Dyson equation, for the static $3D$ photon propagator reads
\begin{align}
  \mathcal{D}_{ij}(\vec q|z,z')= \mathcal{D}_{ij}^0(\vec q|z-z') + e^2\int_{-\infty}^{\infty} \ \mathrm{d}\zeta' \int_{-\infty}^{\infty} \ \mathrm{d}\zeta \;
\mathcal{D}_{ia}^0(\vec q|z-\zeta) \, \delta(\zeta) \sigma_{ab} \delta(\zeta') \mathcal{D}_{bj}(\vec q|\zeta',z') \;,
  \label{eq:63}
\end{align}
where $\vec q$ represents the $2D$ momentum whose magnitude is $\mu\mathfrak{q}$, and all indices, $i,j,a,b = \{x,y\}$ and $e$ is the electric charge in 3+1 dimensions.
The free photon propagator $\mathcal{D}_{ij}^0(q|z-z')$ in this mixed representation (momentum in $x-y$,
coordinate in $z$) takes the form
\begin{align}
  \mathcal{D}_{ij}^0(\vec q|z-z') &= \int_{-\infty}^\infty \frac{\mathrm{d}q_z}{2 \pi} e^{\mathrm{i}q_z(z-z')} \frac{1}{q^2 + q_z^2}\left( \delta_{ij} - \frac{q_iq_j}{q^2 + q_z^2} \right) \; ,
  \label{eq:64}
\end{align}
and the two-dimensional polarization tensor of the medium is written as
\begin{equation}
\sigma_{ab}(\vec q) = \sigma^{\rm tr.}(q)(\delta_{ab} - \frac{q_a q_b}{q^2}) \; .
\end{equation}
For the physical phenomena within the medium, $z=z'=0$ and we have
\begin{align}
  \mathcal{D}_{ij}(\vec q|0,0) &= \mathcal{D}_{ij}^0(\vec q|0) + \int \frac{\mathrm{d}q_z}{2 \pi}  \int \frac{\mathrm{d}q_z'}{2 \pi}
\frac{1}{q^2 + q_z^2} \times \notag \\
  & \hspace{1.5cm}  \left( \delta_{ia} - e^2\frac{q_iq_a}{q^2 + q_z^2} \right)\sigma^{\rm tr.}(q)(\delta_{ab} - \frac{q_aq_b}{q^2})
\frac{1}{q^2 + q_z'^2}\left( \delta_{bj} - \frac{q_bq_j}{q^2 + q_z'^2} \right)  \\
                           &= \mathcal{D}_{ij}^0(\vec q|0) + \frac{1}{4 q^2} \left(\delta_{ij} - \frac{q_iq_j}{q^2} \right) \sigma^{\rm tr.}(q) \; ,
\end{align}
to the order $e^2$. Also, $\mathcal{D}_{ij}(q|0,0)$ is always contracted with the Fourier component of a stationary current in static case and the factor $q_j$ in above equation does not contribute because of the current conservation and hence,  effectively,
$\mathcal{D}_{ij}(\vec q|0,0)=D(q)\delta_{ij}$ with a scalar form factor
\begin{align}
  D(q) = \frac{1}{2 q} + \frac{e^2}{4 q^2} \sigma^{\rm tr.}(q) \; .
  \label{eq:65}
\end{align}
where the first term on RHS comes from the free propagator and the second term reflect the polarization of the medium. For the holographic polarization tensor considered in this work,
$\sigma^{\rm tr.}(q)\propto\mathcal{C}_{yy}(q)$ .
Transforming the second term of Eqn.~(\ref{eq:65}) into coordinate space and denoting the result by $P(\vec{x})$, we have
\begin{align}
  P(\vec x)  \equiv \frac{1}{4} \int \frac{\mathrm{d}^2q}{4\pi^2} e^{\mathrm{i} qx} \frac{\sigma^{\rm tr.}(q)}{q^2} = \frac{1}{8\pi} \int_0^\infty \ \mathrm{d}q \frac{\sigma^{\rm tr.}(q)}{q} J_0(q \cdot |\vec{x}|) \; ,
  \label{eq:66}
\end{align}
where we have used polar coordinates for the momentum integral and $J_0(z)$ is the zeroth order Bessel function.
We employ the technique of contour integral to calculate the radial integral on RHS of Eqn.~(\ref{eq:66}), starting with
\begin{align}
  I \equiv \left (\int_{-\infty + \mathrm{i}0^+}^{0 + \mathrm{i}0^+}+\int_0^{+\infty}\right)
 \mathrm{d}q \,  \frac{\sigma^{\rm tr.}(q)}{q} H_0^{(1)}( q \cdot |\vec{x}|) \, ,
  \label{eq:67}
\end{align}
where $H_0^{(1)}(z)=J_0(z)+\mathrm{i} Y_0(z)$ is the zeroth order Hankel function of the first kind and $Y_0(z)$ is the zeroth order Neumann function,
\begin{align}
  Y_0(z) = \frac{2}{\pi}J_0(z)\ln \frac{z}{2} + \text{ an analytic function even w.r.t } z \; .
  \label{eq:68}
\end{align}
The integration path of (\ref{eq:67}) is chosen to run just above the logarithmic cut along the negative real axis.
Since $J_0(q\cdot|\vec{x}|)$ and $\sigma^{\rm tr.}(q)$ are even with respect to $q$, the non-zero result of $I$ is given by the non-even part of $H_0^{(1)}( q \cdot |\vec{x}|)$
with respect to $q$, and reads
\begin{align}
  I = - 2 \int_{-\infty}^0 \ \mathrm{d}q \frac{\sigma^{\rm tr.}(q)}{q} J_0(q\cdot|\vec{x}|)
=- 2 \int_0^{+\infty} \ \mathrm{d}q \frac{\sigma^{\rm tr.}(q)}{q} J_0(q\cdot|\vec{x}|) = -16\pi P(\vec{x})\; .
  \label{eq:69}
\end{align}
To calculate the integral in (\ref{eq:67}), we assume that there are no poles on the entire physical sheet and deform the contour on the upper-half $q$-plane to wrap up the pair of branch-cuts on the right panel of Fig.~\ref{fig: comparison}. We obtain that $P(\vec{x}) =\rm{Re}\, I = \rm{Re} \, (I_+ + I_-)$ with
\begin{align}
  I_\pm = \oint_{C_\pm}\frac{\sigma^{\rm tr.}(q)}{q} H_0^{(1)}(q\cdot|\vec{x}|) \ \mathrm{d}q \; ,
\label{eq:72}
\end{align}
where $C_\pm$ denotes a contour wrapping up the cut originated from the branch point $q_\pm=\left(\pm\frac{1}{2}+\mathrm{i} \right)\mu$ and turning around the branch point counterclockwisely.
For large $|\vec x|$
\begin{align}
  H_0^{(1)}(q\cdot|\vec{x}|) \sim \sqrt{\frac{2}{\pi q\cdot|\vec{x}|}} e^{\mathrm{i} (q\cdot|\vec{x}| - \frac{\pi}{4})} \; ,
  \label{eq:73}
\end{align}
we have
\begin{align}
  I_\pm = \sqrt{\frac{2}{\pi|\vec{x}|}}e^{-\mathrm{i}\frac{\pi}{4}}\oint_{C_\pm}\frac{\sigma^{\rm tr.}(q)}{q^{3/2}}e^{ \mathrm{i} q\cdot|\vec x|} \ \mathrm{d}q \; .
\label{eq:72}
\end{align}
According to the analysis in section \ref{sec:analyt-magn-susc}, the branch points are of square root type and $\sigma^{\rm tr.}(q)$ does not diverging at the branch points. Hence we may write
\begin{equation}
\frac{\sigma^{\rm tr.}(q)}{q^{3/2}}=f(q)+g(q)\sqrt{q-q_\pm}
\end{equation}
and the functions $f(q)$ and $g(q)$ can be expanded according to integer powers of $t\equiv q-q_\pm$. The term that dominates the large $|\vec x|$ behavior is the leading term of the power series of $g(q)$, which predominately contributes to the contour integral, i.e.
\begin{align}
  I_\pm \simeq \sqrt{\frac{2}{\pi|\vec{x}|}}e^{-i\frac{\pi}{4}}g(q_\pm)e^{\pm iq_\pm|\vec x|}\oint_{C_\pm}t^{1/2}e^{i|\vec x|t} \ \mathrm{d}t
\sim \frac{1}{|\vec x|^2}e^{\mathrm{i}q_\pm|\vec x|} \; ,
\label{eq:72}
\end{align}
Consequently,
\begin{equation}
P(\vec x) = \mathrm{Re} \, \{ I_+ + I_- \} \sim \frac{1}{|\vec x|^2}e^{-\mu|\vec x|}\cos\left(\frac{\mu}{\sqrt{2}}|\vec x|+\phi\right) \, ,
\end{equation}
where the phase $\phi$ depends on the phases of $I_+$ and $I_-$. The large-$|\vec{x}|$ behavior (\ref{eq:59}) is thereby obtained.


\providecommand{\href}[2]{#2}\begingroup\raggedright\endgroup

\end{document}